\documentclass{camera}
\usepackage{graphicx}  
\usepackage{amsmath,amssymb}

\begin{document}

%
\title{Symmetry energy in the structure\\ and in reactions}

%
\author{P. Danielewicz}

%
\organization{National Superconducting Cyclotron Laboratory\\ and Department of Physics and Astronomy\\ Michigan State University, East Lansing, MI 48824-1321, USA}

\maketitle

\begin{abstract}
Efforts to extract information on magnitude and density dependence of the nuclear symmetry energy are discussed.  The utilized data include those on mass dependence of the excitation energies to the isobaric analog states of ground states, as well as data on the diffusion of isospin in heavy-ion reactions.  Results following from different observables are compared.
\end{abstract}

%
The interest in nuclear symmetry energy has grown on account of the relatively recent availability of exotic beams, allowing to study systems with an increased range of relative asymmetry, $\eta = (N-Z)/A$, for a given mass number $A$.  Investigations have progressed focusing alternatively on the structure of nuclei, processes taking place in reactions and on collective excitations.  Constraints on the symmetry energy would improve extrapolations from finite nuclei to neutron stars.

Understanding of structural features, associated with the nuclear symmetry energy, can be advanced following the symmetries of nuclear interactions: charge symmetry, which is the symmetry under neutron-proton exchange, and charge invariance, which is the symmetry under rotations in neutron-proton space.  In the context of charge symmetry, isoscalar quantities can be introduced, that do not change under neutron-proton interchange.  If an isoscalar quantity $F$ is expanded in $\eta$, the expansion contains even powers only:
\begin{equation}
F(\eta) = F_0 + F_2 \, \eta^2 + F_4 \, \eta^4 + \ldots \, .
\end{equation}
Because of the lack of a linear term and because of small $\eta$-values in nuclei, $\eta \lesssim 1/4$, isoscalar quantities depend weakly on asymmetry.  Examples of isoscalar quantities include nuclear energy and net nuclear density $\rho=\rho_n+\rho_p$.  Isovector quantities are those that change sign under neutron-proton interchange.  Example of isovector quantity is the neutron-proton density difference, $\rho_{np} = \rho_n - \rho_p$.  An isovector quantity, expanded in $\eta$, contains odd powers only:
\begin{equation}
G(\eta) = G_1 \, \eta + G_3 \, \eta^3 + \ldots \, .
\end{equation}
Notably, an isovector quantity divided by the asymmetry, $G/\eta$, or by another isovector quantity, becomes an isoscalar quantity and, in this, weakly dependent on $\eta$.

The considerations above apply to quantities with a continuous dependence on $\eta$.  Microscopic, shell and pairing, effects can introduce discontinuous changes.  However, the considerations with expansion should apply to quantities that are averaged over the microscopic effects.  In the context of charge invariance, the isoscalar quantities are those that do not change under rotations in neutron-proton space and the isovector quantities are those that rotate in a covariant manner.  One consequence of the charge invariance of nuclear interactions is the appearance of isobaric analog multiplets across the isobaric chains, at about the same energy.  Coulomb interactions break either symmetry, but may be accounted for in terms of correction terms.

Up to the second order in $\eta$, the nuclear contribution to the nuclear energy may be represented as
\begin{equation}
\label{eq:ENZ}
E(N,Z) = E_0(A) + \frac{a_a(A)}{A} \, (N-Z)^2 \, .
\end{equation}
In simple mass formulas, the symmetry coefficient $a_a$ is usually assumed to be constant.  However, the symmetry considerations alone do not prevent this coefficient from being $A$-dependent, which we shall retain for generality.  In \eqref{eq:ENZ}, we may note a similarity to the energy of a capacitor in electrostatics, with capacitance $C$ and  charge $Q$,
\begin{equation}
E=E_0 + \frac{Q^2}{2 C} \, ,
\end{equation}
where $E_0$ is the energy of the capacitor without charge.  We can recognize that the asymmetry $N -Z$ corresponds to the charge $Q$ and $A/(2 a_a)$ corresponds to the capacitance $C$.  The analog of the capacitor voltage,
\begin{equation}
V = \frac{\partial E}{\partial Q} = \frac{Q}{C} \, ,
\end{equation}
is the asymmetric chemical potential
\begin{equation}
\mu_a = \frac{\partial E}{\partial (N - Z)} = \frac{2 a_a(A)}{A} \,(N - Z) \, ,
\end{equation}
equal to, as in electrostatics, to the asymmetry charge divided by the capacitance for asymmetry.

Consistently with charge symmetry, the energy per nucleon in uniform matter may be represented, for low $\eta$, as
\begin{equation}
\label{eq:E0S}
\frac{E}{A} ( \rho_n, \rho_p) = \frac{E_0}{A} (\rho) + S(\rho) \, \eta^2 \, .
\end{equation}
However, microscopic calculations, such as~\cite{bom91}, indicate that the r.h.s.\ of \eqref{eq:E0S} represents the l.h.s.\ rather accurately all the way up to $|\eta| = 1$, for a wide range of $\rho$.  In consequence, the two functions of net density, $\frac{E_0}{A} (\rho)$ and $S(\rho)$, are sufficient to describe accurately the energy in uniform matter at different combinations of $\rho_n$ and $\rho_p$.  There is an interest in the expansion of those functions around the normal density $\rho_0$.  By definition, $E_0/A$ minimizes at $\rho_0$, but $S$ generally has a finite slope, typically quantified in terms of the constant $L$:
\begin{equation}
S(\rho) = a_a^V + \frac{L}{3} \frac{\rho - \rho_0}{\rho_0} + \ldots \, .
\end{equation}
Here, $a_a^V = S(\rho_0)$ represents the symmetry coefficient for a large system dominated by normal density.
Because of the minimum of $E_0/A$, the contribution from symmetry energy in \eqref{eq:E0S} tends to dominate the pressure in neutron stars, at densities of the order of normal.

As has been discussed, the net density, $\rho = \rho_n - \rho_p$, is isoscalar and, thus, should be weakly dependent on $\eta$, for a given~$A$.  We commonly parameterize $\rho$ as
\begin{equation}
\rho(r) = \frac{\rho_0}{1 + \text{exp} \left( \frac{r-R}{d} \right)} \, ,
\end{equation}
with $R= r_0 \, A^{1/3}$.
On the other hand, the density difference, $\rho_n - \rho_p$, is isovector.  However, the ratio $(\rho_n - \rho_p)/\eta$ is isocalar.  For a nucleus, the asymmetry $\eta$ in the latter normalization is a global quantity and it turns out to be more convenient to normalize the density difference in terms of the intense asymmetric chemical potential, to yield the asymmetric density:
\begin{equation}
\rho_a (r) = \frac{2 a_a^V}{\mu_a} \left[\rho_n(r) - \rho_p(r) \right] \, .
\end{equation}
In the limit of a large system dominated by $\rho_0$, the two normalizations yield the same result.  The two densities, $\rho$ and $\rho_a$ both weakly depend on $\eta$.  Further, as will become partially apparent, the density $\rho_a$ is universally related to~$\rho$.  Out of those two densities, the densities of neutrons and protons in different nuclei may be constructed:
\begin{equation}
{\rho_{n,p}(r) = \frac{1}{2} \big[ {\rho(r)} \pm \frac{\mu_a}{2 a_a^V} { \rho_a(r) } \big]} \, .
\end{equation}

The asymmetric density, that represents a formfactor for the isovector difference $(\rho_n - \rho_p)$, is related, on one hand, to the generalized symmetry coefficient $a_a(A)$ and, on the other, to a local value of the symmetry energy $S(\rho)$.  The first relation follows from the fact that the capacitance for asymmetry may be represented as the ratio of asymmetry (charge in the electrostatic analogy) to the asymmetric potential (voltage), yielding
\begin{equation}
\frac{A}{a_a(A)} = \frac{2(N-Z)}{\mu_a} = 2 \int \text{d}r \, \frac{\rho_{np}}{\mu_a}
= \frac{1}{a_a^V} \int \text{d}r \, {\rho_a(r)} \, .
\end{equation}
The integral over $\rho_a$ is thus proportional to the capacitance for asymmetry.  The farther the asymmetric density sticks out from the nuclear volume, the greater the capacitance for asymmetry.  Otherwise, in uniform matter, we find from the definition of the chemical potential
\begin{equation}
\mu_a = \frac{2 S(\rho)}{\rho} \, (\rho_n - \rho_p) \, ,
\end{equation}
which yields for the asymmetric density
\begin{equation}
\label{eq:rhoaS}
\rho_a = \frac{a_a^V \, \rho}{S(\rho)} \, .
\end{equation}
Due to the short-range of nuclear interactions, the result \eqref{eq:rhoaS} is further expected to be approximately valid in weakly nonuniform matter.

\begin{figure}
\includegraphics[width=.49\linewidth,height=.62\linewidth]{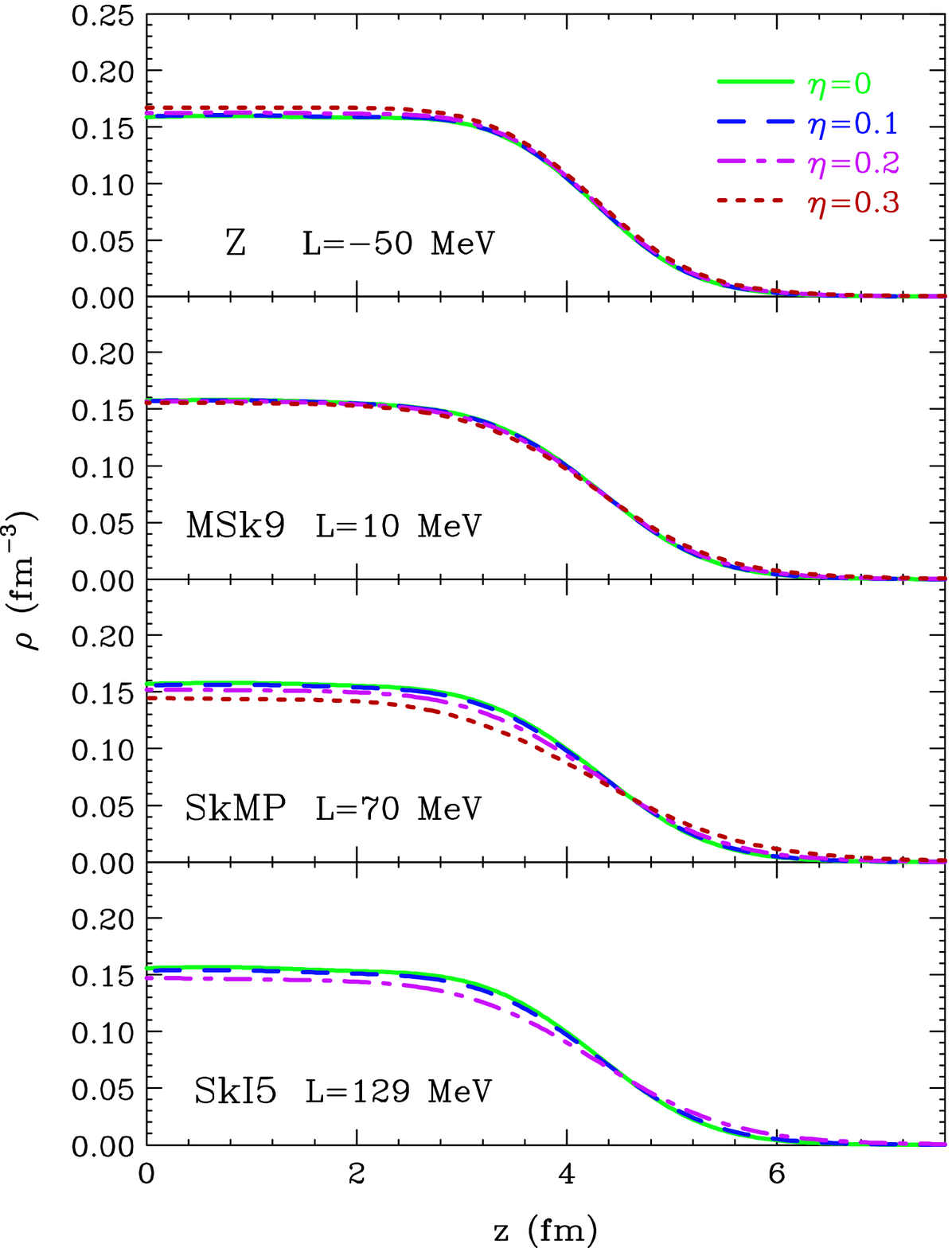}
\hfill
\includegraphics[width=.49\linewidth,height=.62\linewidth]{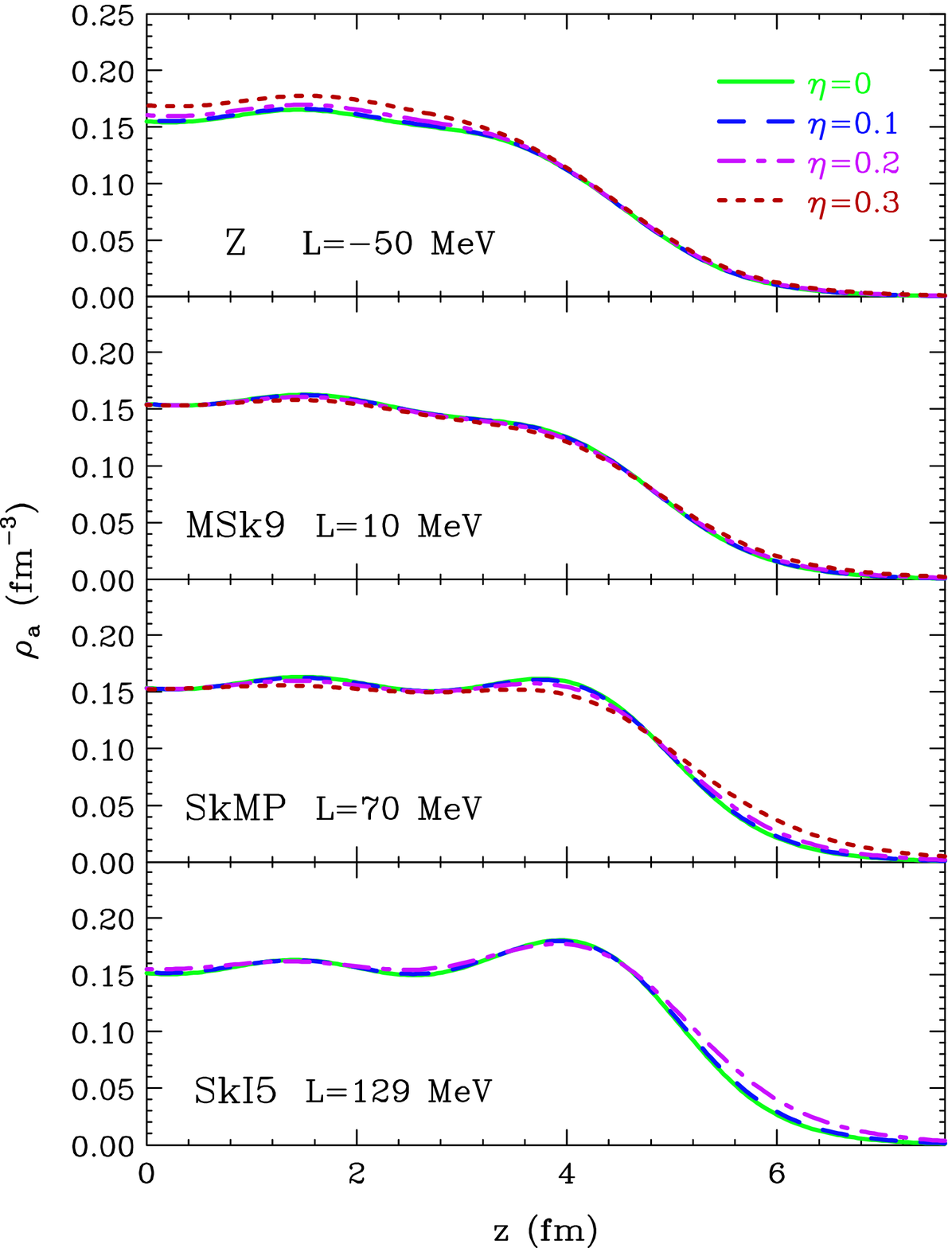}
\caption{Net (isoscalar, on the left) and isovector (right) densities at different asymmetries in half-infinite nuclear matter, as a function of position, for different Skyrme interactions, after Ref.~\cite{dan09}.
}
\label{fig:anpr01} 
\end{figure}

The anticipated weak dependence of the two densities on asymmetry is next tested in Fig.~\ref{fig:anpr01}.  The separate panels show $\rho$ (left) and $\rho_a$ (right), for different Skyrme interactions (from top to bottom), in Hartree-Fock calculations of half-infinite nuclear matter~\cite{dan09}, at different asymmetries (different lines).  In half-infinite matter, the shell effects are suppressed.  In addition, the Coulomb interactions are switched off, eliminating the need for any Coulomb corrections.  It is evident in Fig.~\ref{fig:anpr01} that the two densities indeed change very little with the asymmetry.

\begin{figure}
\includegraphics[width=.54\linewidth,height=.50\linewidth]{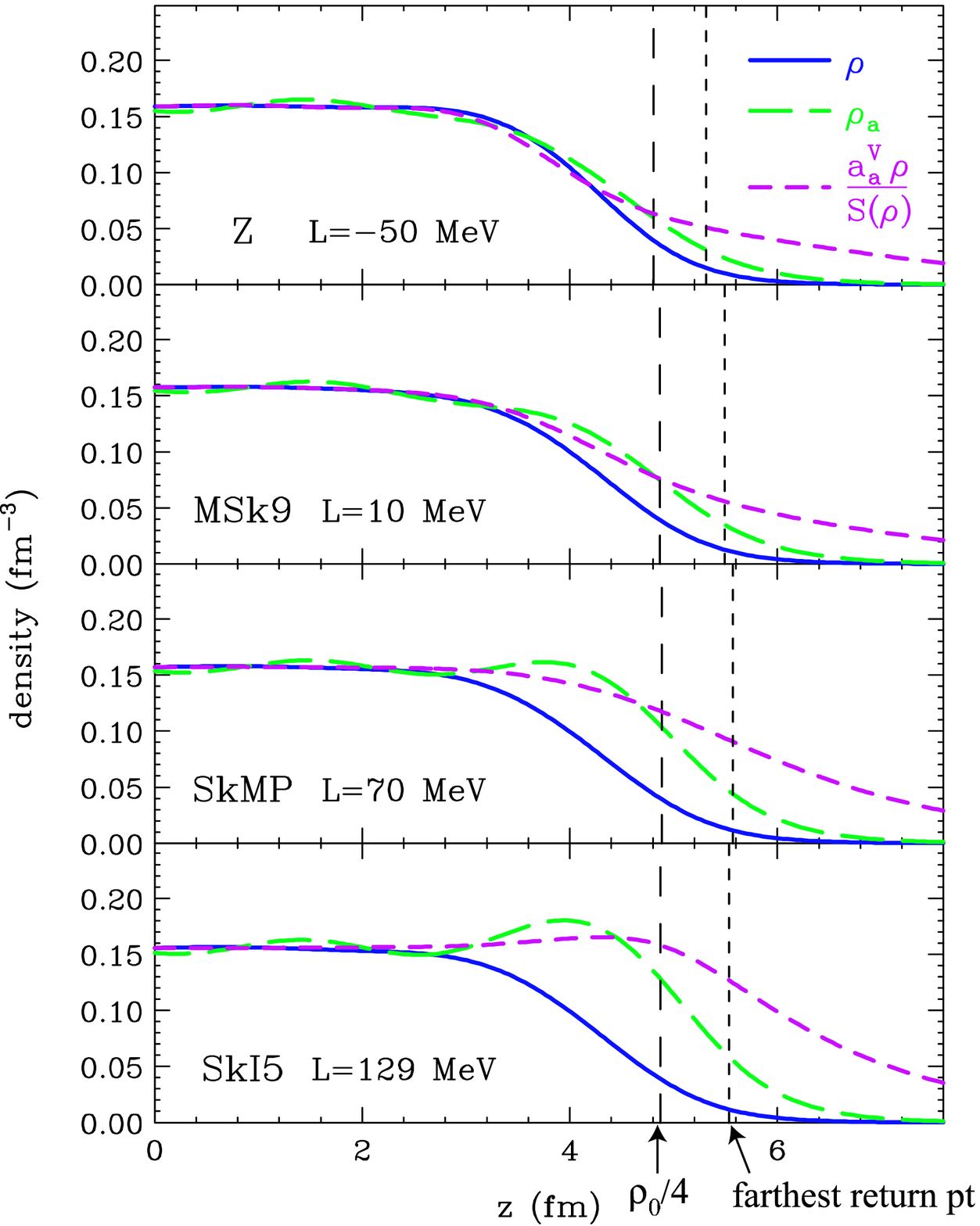}
\hfill
\includegraphics[width=.44\linewidth,height=.40\linewidth]{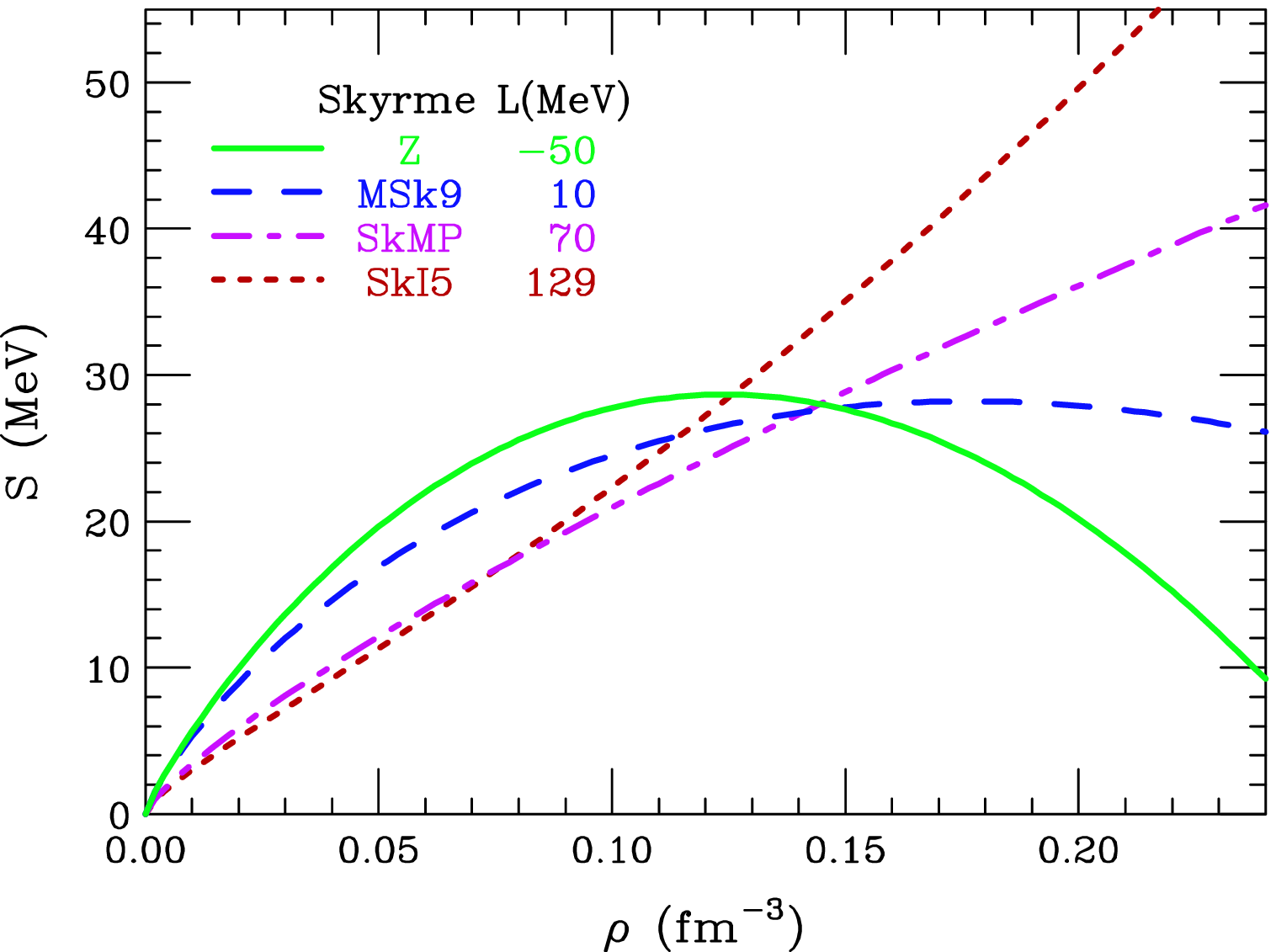}
\caption{The left set of panels shows different densities in half-infinite nuclear-matter, as a function of position, for different Skyrme interactions, after Ref.~\cite{dan09}.  The solid curves there represent the net density at asymmetry $\eta=0$.  The short-dashed curves represent the isovector density at $\eta=0$.  Finally, the long-dashed curves represent expectations for the isovector density based on the local value of the symmetry energy.  The right panel shows the dependence of symmetry energy on density, for different Skyrme interactions.}
\label{fig:anpros} 
\end{figure}

Left panels of Fig.~\ref{fig:anpros} next compare the isoscalar and isovector densities for the different interactions.  Within the matter, the densities are fairly close to each other.  However, in the surface area differences emerge that are strongly correlated to the $L$-value associated with the symmetry energy.  The higher the $L$-value, the farther out is the isovector density displaced relative to the isoscalar density.  This can be understood in terms of the behavior of the symmetry energy at low densities, seen in the right panel of Fig.~\ref{fig:anpros}.  The isovector density $\rho_a$ follows the expectation of Eq.~\eqref{eq:rhoaS} down to about the density of $\rho \approx \rho_0/4$ a representative classical return point.  Low values of $S(\rho)$, at $\rho \lesssim \rho_0$, enhance $\rho_a$, as evident in Fig.~\ref{fig:anpros}.  Densities $\rho \lesssim \rho_0/4$ are dominated by tunneling and local relations are not expected to hold.

In the context of surface differences, for a large system we can get for the capacitance for asymmetry:
\begin{equation}
\begin{split}
2 C \equiv \frac{A}{a_a(A)} =  \frac{1}{a_a^V} \int \text{d}^3 r \, \rho_a (r)
&  = \frac{1}{a_a^V} \int \text{d}^3 r \, \rho (r) + \frac{1}{a_a^V} \int \text{d}^3 r \, (\rho_a - \rho) (r)\\
&  \simeq \frac{A}{a_a^V} + \frac{A^{2/3}}{a_a^S} \, .
\label{eq:2C}
\end{split}
\end{equation}
The last approximate equality follows from the fact that the two densities are substantially different only in the surface region.
We see here that the capacitance emerges here as a sum of two capacitances, one associated with nuclear interior, proportional to~$A$, and one associated with the surface, proportional to~$A^{2/3}$. The surface capacitance, in terms of $a_a^S$, is tightly correlated with the slope-parameter~$L$.  Low values of $L$ are associated with large surface capacitance and low values of~$a_a^S$.
Even tighter, and better physically justified, is the correlation between the ratios $a_a^V/a_a^S$ and $L/a_a^V$~\cite{dan09}.  Given the degree to which which Eq.~\eqref{eq:rhoaS} is followed for different interactions, the latter correlation is expected to be robust.

Determination of the asymmetry coefficient with its mass dependence, by fitting nuclear masses with an energy formula, is difficult because the symmetry-energy contribution is small and its details compete against details of other contributions to the energy~\cite{dan03}.  However, upon generalizing the symmetry-energy term in an energy formula, effects of the symmetry term may be studied on a nucleus by nucleus basis, in isolation from other contributions to the energy of a nucleus.  Specifically, we can observe that the symmetry energy may be represented in terms of the isospin of the nucleus $(T,T_z)$ as
\begin{equation}
E_a = a_a(A) \, \frac{(N-Z)^2}{A} = 4 \, a_a(A) \,  \frac{T_z^2}{A} \, .
\end{equation}
This representation makes it apparent that, in the present form, the symmetry energy is an isoscalar under charge symmetry, but not under charge invariance.  However, if we replace the square of the third component of isospin with the square of the net isospin, we will arrive at an isoscalar under charge invariance, required for the energy under that symmetry,
\begin{equation}
E_a = 4 \, a_a(A) \, \frac{T^2}{A}= 4 \, a_a(A) \, \frac{T(T+1)}{A} \, .
\end{equation}
This result should apply to a lowest state with a given isospin in the nucleus.  Such excited states represent isobaric analog states (IAS) of the ground states of nuclei with a higher asymmetry in the specific isobaric chain.  In the ground state, the quantum number for the net isospin agrees in magnitude with the third component, $T=|T_z|$.

\begin{figure}
\centerline{\includegraphics[width=.74\linewidth]{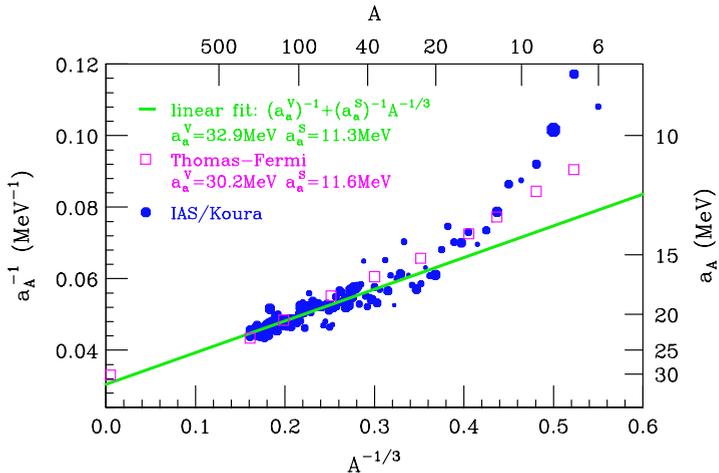}}
\caption{Inverse values of symmetry coefficients plotted plotted vs inverse values of the cube root of mass number.  Circles represent values derived from excitation energies to the states that are isobaric analogs of ground states \cite{ant97}, when utilizing shell corrections by Koura {\em et al.} \cite{kou05}.  Line represents a linear fit to the results from analog states.  Squares represent results from a Thomas-Fermi calculation~\cite{dan03}.}
\label{fig:Inverseaa} 
\end{figure}

With the generalization, the excitation energy to an IAS becomes
\begin{equation}
\begin{split}
E_2(T_2)-E_1(T_1) = & \frac{4 \, a_a}{A} \big\lbrace T_2(T_2+1) - T_1 (T_1+1) \big\rbrace \\
& { + E_\text{mic}(T_2,T_z) - E_\text{mic}(T_2,T_z)} \, ,
\end{split}
\end{equation}
where we also account for corrections due to microscopic effects and deformation.  We employ the corrections by Koura {\em et al.}~\cite{kou05} and obtain generalized symmetry coefficients on a nucleus-by-nucleus basis from
\begin{equation}
a_a = \frac{A \, \Delta E}{4 \Delta T^2} \, ,
\end{equation}
using data on IAS compiled in Ref.~\cite{ant97}.
Inverse values of the coefficient are plotted as a function of $A^{-1/3}$ in Fig.~\ref{fig:Inverseaa}. The coefficient values drop at low $A$, down to $\sim 10 \, \text{MeV}$, which corresponds to an increase in capacitance per nucleon compared to heavy nuclei.  In heavy nuclei, the coefficient values rise up to $\sim 23 \, \text{MeV}$.

Figure~\ref{fig:Inverseaa} demonstrates that the mass dependence of the symmetry coefficient may be fitted, at larger $A$, with
\begin{equation}
a_a^{-1}(A) = (a_a^V)^{-1} + (a_a^S)^{-1} \, A^{-1/3} \, ,
\label{eq:aaVS}
\end{equation}
which follows from Eq.~\eqref{eq:2C}.  This, principally, allows for a model-independent determination of the coefficients $a_a^V$ and $a_a^S$.  With the correlation between $a_a^V/a_a^S$ and $L/a_a^V$, further, the value of $L$ may be estimated.  Figure~\ref{fig:Inverseaa} shows also results from the simple Thomas-Fermi theory~\cite{dan03} that has the benefit of producing some curvature effect.  Progressing in such a fashion, one can estimate the nuclear values of interest: $a_a^V = (31.5-33.5) \, \text{MeV}$, $a_a^S = (9.5-12) \, \text{MeV}$ and $L \sim 95 \, \text{MeV}$~\cite{dan09}.

\begin{figure}
\includegraphics[width=.48\linewidth]{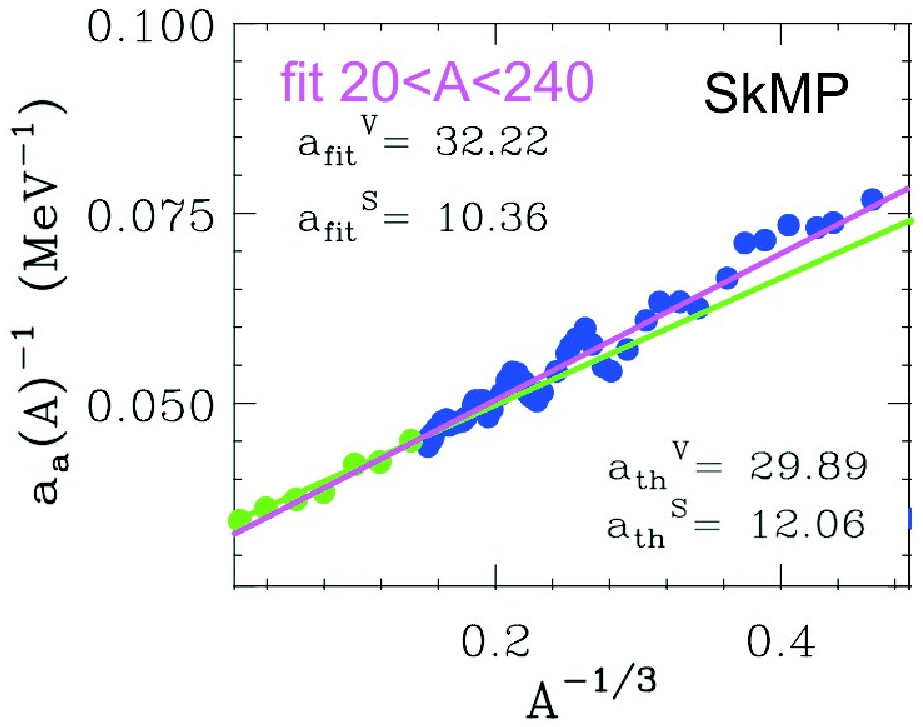} \hfill \includegraphics[width=.48\linewidth]{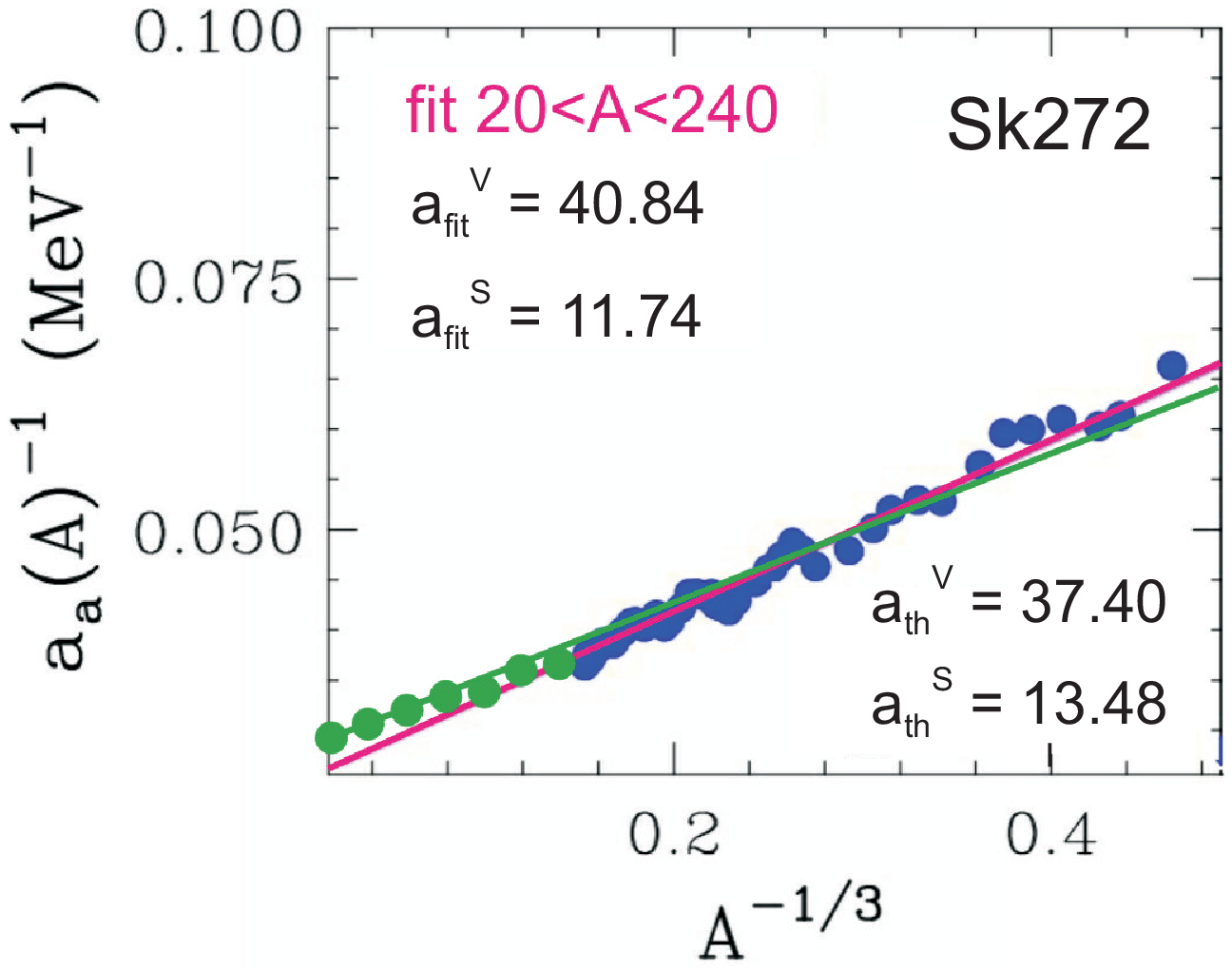}
\caption{Inverse values of symmetry coefficients, for the SkMP and SK272 interactions, plotted vs inverse values of the cube root of mass number.  Symbols represent results from spherical Skyrme-Hartree-Fock calculations using the codes by Reinhard for realistic~\cite{rein91} and unrealistically large~\cite{rein06} nuclei with Coulomb interactions switched off.  The lines represent, respectively, linear fits to nuclei in the mass region $20 < A < 240$ and expectations from calculations of half-infinite matter~\cite{dan09}.}
\label{fig:SkMP} 
\end{figure}

\begin{figure}
\centerline{\includegraphics[width=.54\linewidth]{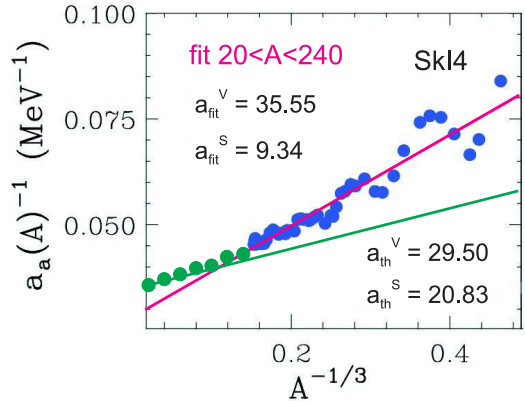}}
\caption{Inverse values of symmetry coefficients, for the SkI4 interaction, plotted vs inverse values of the cube root of mass number.  Symbols represent results from spherical Skyrme-Hartree-Fock calculations using the codes by Reinhard for realistic~\cite{rein91} and unrealistically large~\cite{rein06} nuclei with Coulomb interactions switched off.  The lines represent, respectively, a linear fit to nuclei in the mass region $20 < A < 240$ and expectation from calculations of half-infinite matter~\cite{dan09}.}
\label{fig:SkI4} 
\end{figure}

\begin{figure}
\includegraphics[width=.50\linewidth]{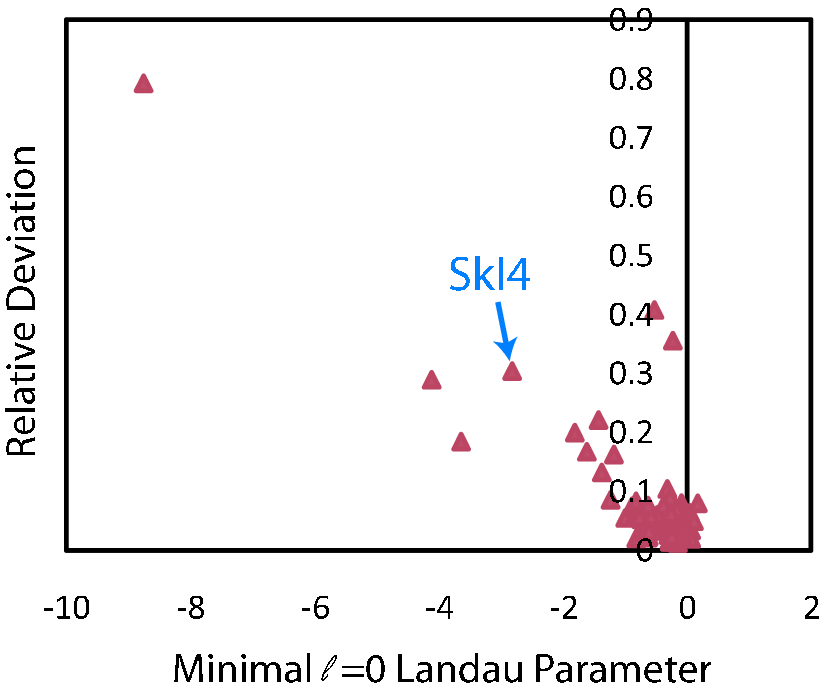}
\hfill
\includegraphics[width=.50\linewidth]{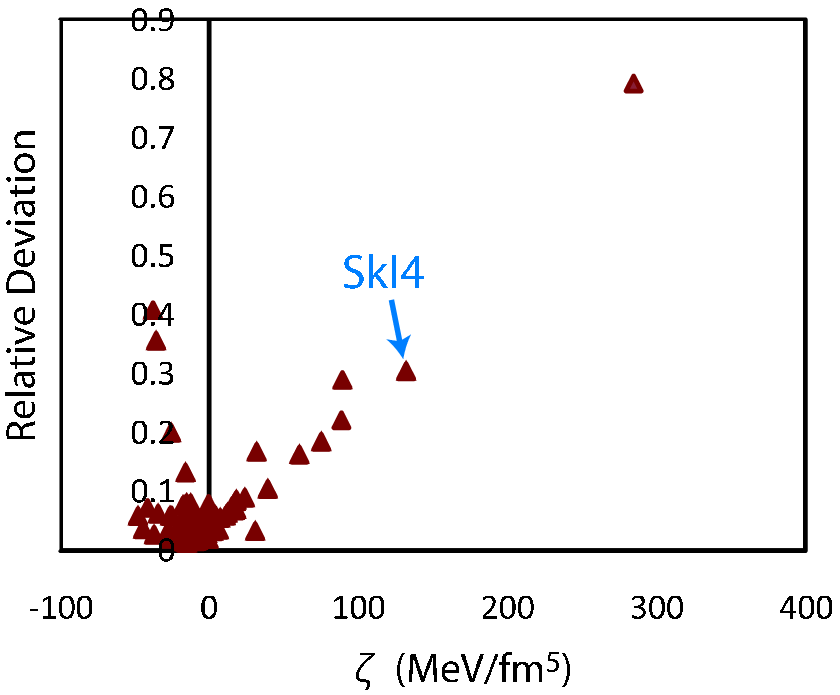}
\caption{Relative rms deviation of the mass-dependent symmetry coefficients, expected from calculations of semi-infinite matter, from coefficients from spherical calculations, for a variety of Skyrme interactions, plotted vs a minimal $\ell =0$ Landau coefficient (left panel) and vs the $\zeta$-parameter for the Skyrme energy functional.}
\label{fig:LandauZee} 
\end{figure}

For the results such as above to be robust, though, the results following from an analysis need to agree with those underlying the theory, when realistic description of nuclei is applied.  This is next tested in Fig.~\ref{fig:SkMP}, which shows sample results of calculations~\cite{dan10} of the symmetry coefficients within the Skyrme-Hartree-Fock calculations of unrealistically large~\cite{rein06} and realistic~\cite{rein91} masses, together with expectations based on the results for half-infinite matter~\cite{dan09}.  When the formula \eqref{eq:aaVS} is fitted to the medium and heavy nuclei found in nature, typically the coefficients result which are close to those established for the interaction in the infinite and half-infinite calculations.  There are, however, some systematic differences.  Namely, the fitted $a_a^V$ values tends to be a bit higher and $a_a^S$ values a bit lower than in the direct calculations.  With this, the fitted values should lead to somewhat excessive $L$-values.  For a few of the Skyrme interactions \cite{dan09}, however, we find large differences between the fitted and underlying values, such as for the SkI4 interaction in Fig.~\ref{fig:SkI4}.  The situation is further illustrated in Fig.~\ref{fig:LandauZee} that shows the relative deviations between the fitted and expected linear dependencies of $1/a_a$ on $A^{-1/3}$, over $20 < A < 250$ mass region, for about 150 different Skyrme parameterizations from the literature~\cite{dan09,dan10}.  For the majority of those interactions the deviations are small, but for a small percentage of those interactions the deviations are large.  The interactions with the large deviations tend to have objectively unphysical features such as be unstable in the long-wavelength limit or exhibit unphysically strong nonlocality in the symmetry energy.  Long-wavelength instabilities are signalled by one of the $\ell=0$ Landau coefficients being lower than $-1$.  Correspondingly, in the left panel of Fig.~\ref{fig:LandauZee} the deviations between the fit and expectations are plotted against the lowest of the $\ell=0$ Landau coefficients.  It is seen that, indeed, large negative values of the lowest $\ell=0$ coefficient are associated with large deviations.  Contribution of interactions to the nonlocality of symmetry energy, for Skyrme parameterizations, may be quantified in terms of the so-called coefficient $\zeta$ in the term within the Skyrme Hamiltonian density~\cite{ton84}
\begin{equation}
\zeta \, \left[ \nabla (\rho_n - \rho_p) \right]^2 \, .
\end{equation}
Excessively large magnitudes of $\zeta$ represent senselessly long range of inter-nucleon interactions.  In the right panel of Fig.~\ref{fig:LandauZee}, the deviations are plotted against the $\zeta$-values.  Again here it is seen that large $\zeta$-values are associated with large deviations. It is apparent that before reliable corrections for curvature may be established, and firmer conclusions may be reached on $a_a^V$, $a_a^S$ and $L$, associated with the limits of infinite and half-infinite matter, the Skyrme interactions need to be filtered to reject those which exhibit significant nonphysical features of one type or another.

\begin{figure}
\includegraphics[width=.40\linewidth]{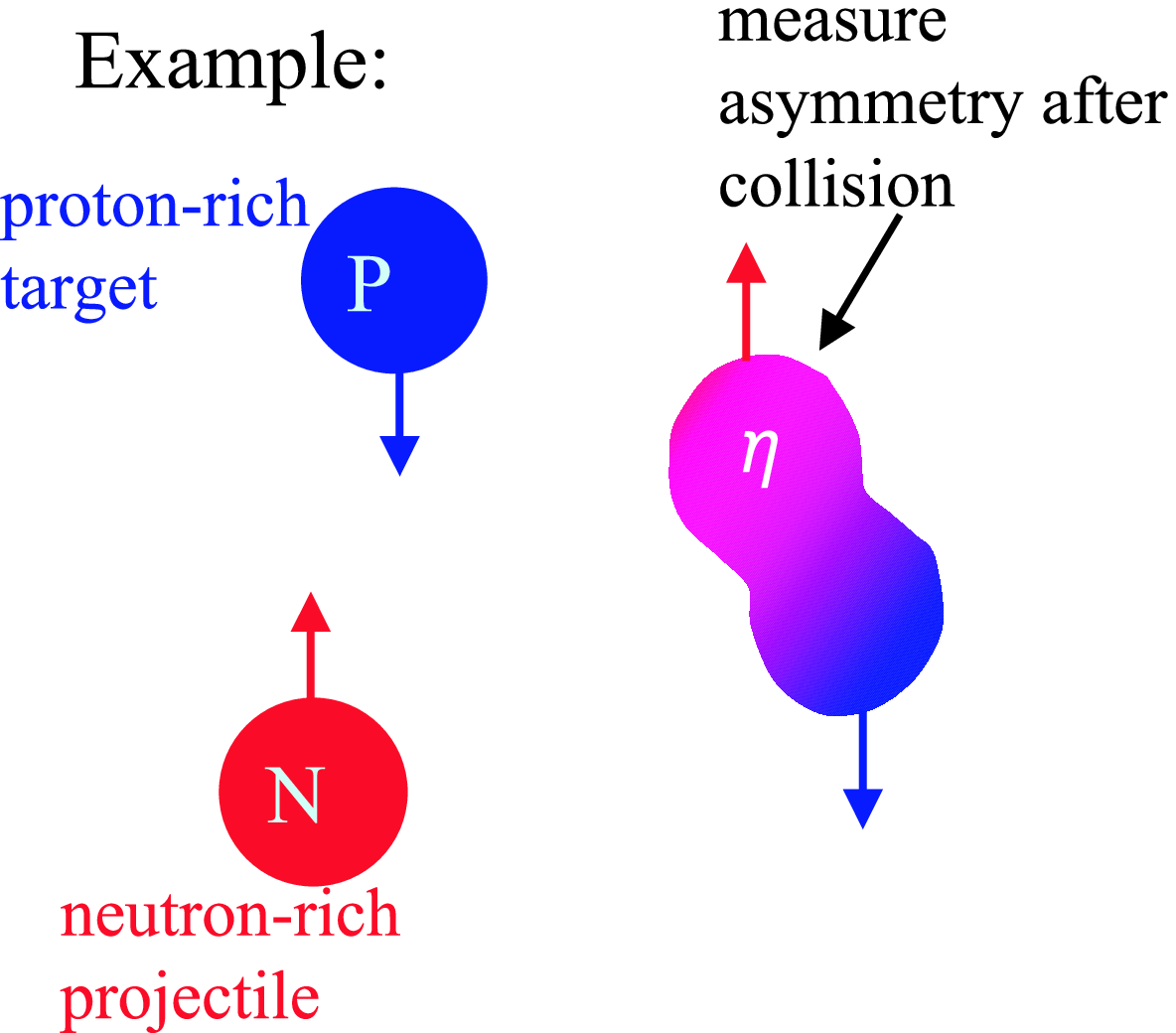}
\hfill
\includegraphics[width=.60\linewidth]{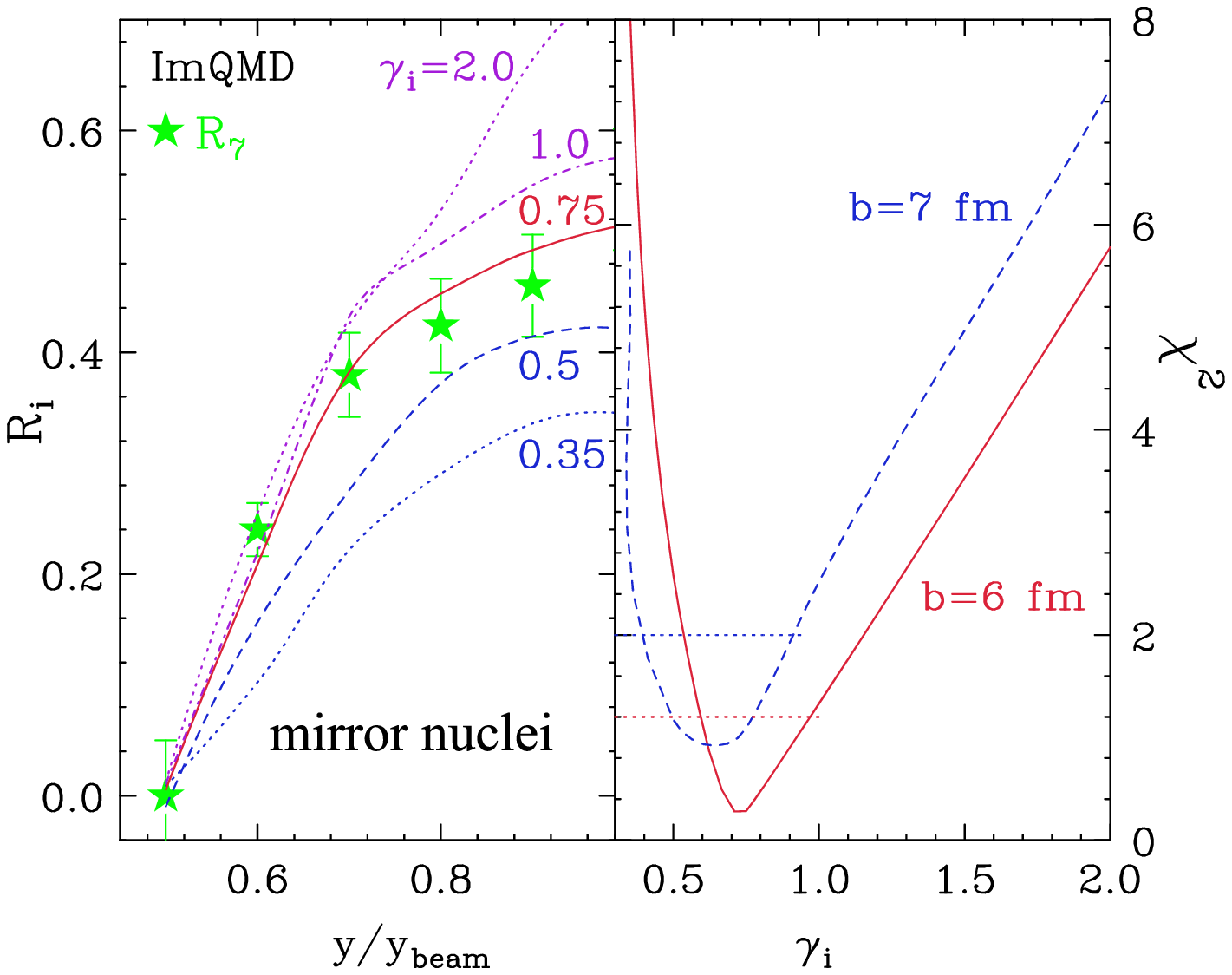}
\caption{Stronger variation of symmetry energy with position enhances transport of neutron-proton asymmetry across the reacting system, sketched on the left.  On the right, the results of transport calculations with different parameterizations of the symmetry energy are compared to data~\cite{liu07} on isospin equilibration, from Sn + Sn collisions at 50~MeV/nucleon, after Ref.~\cite{tsa09}.}
\label{fig:IR_bet09} 
\end{figure}

Within central nuclear reactions, the symmetry energy and its density dependence can be assessed by examining the transport of neutron-proton asymmetry, or isospin, across the reacting system~\cite{tsa04,gal09}, see Fig.~\ref{fig:IR_bet09}.  Changing symmetry energy can assist in the transport of isospin across the reaction zone~\cite{shi03,bar05}, just as electric field can assist in the transport of charge along a wire.  To minimize effects of the loss of isospin to the outside of the system, rather than transport across, different combinations of nuclei may be studied in an experiment, symmetric and asymmetric~\cite{ram00,tsa04,tsa09}, such as the neutron-rich $^{124}$Sn + $^{124}$Sn, more proton-rich $^{112}$Sn + $^{112}$Sn and mixed $^{124}$Sn + $^{112}$Sn.  The variable, which emphasizes the effects of isospin transport and facilitates comparisons between experiment and theory, is the ratio~\cite{ram00}
\begin{equation}
R = 2 \, \frac{\eta_\text{mixed}  - \frac{1}{2} (\eta_\text{n-rich} + \eta_\text{p-rich})}{\eta_\text{n-rich} - \eta_\text{p-rich}} \, .
\end{equation}
In the absence of isospin transport, the ratio $R$ should reach values of $\pm 1$ in the projectile and target regions. In the case of complete isospin mixing, the expected ratio is $R \approx 0$ across the system.  The convenience of the ratio in comparisons is that the asymmetry may be replaced by any quantity expected to be proportional to the asymmetry, with expected similar results for $R$.

\begin{figure}
\includegraphics[width=\linewidth]{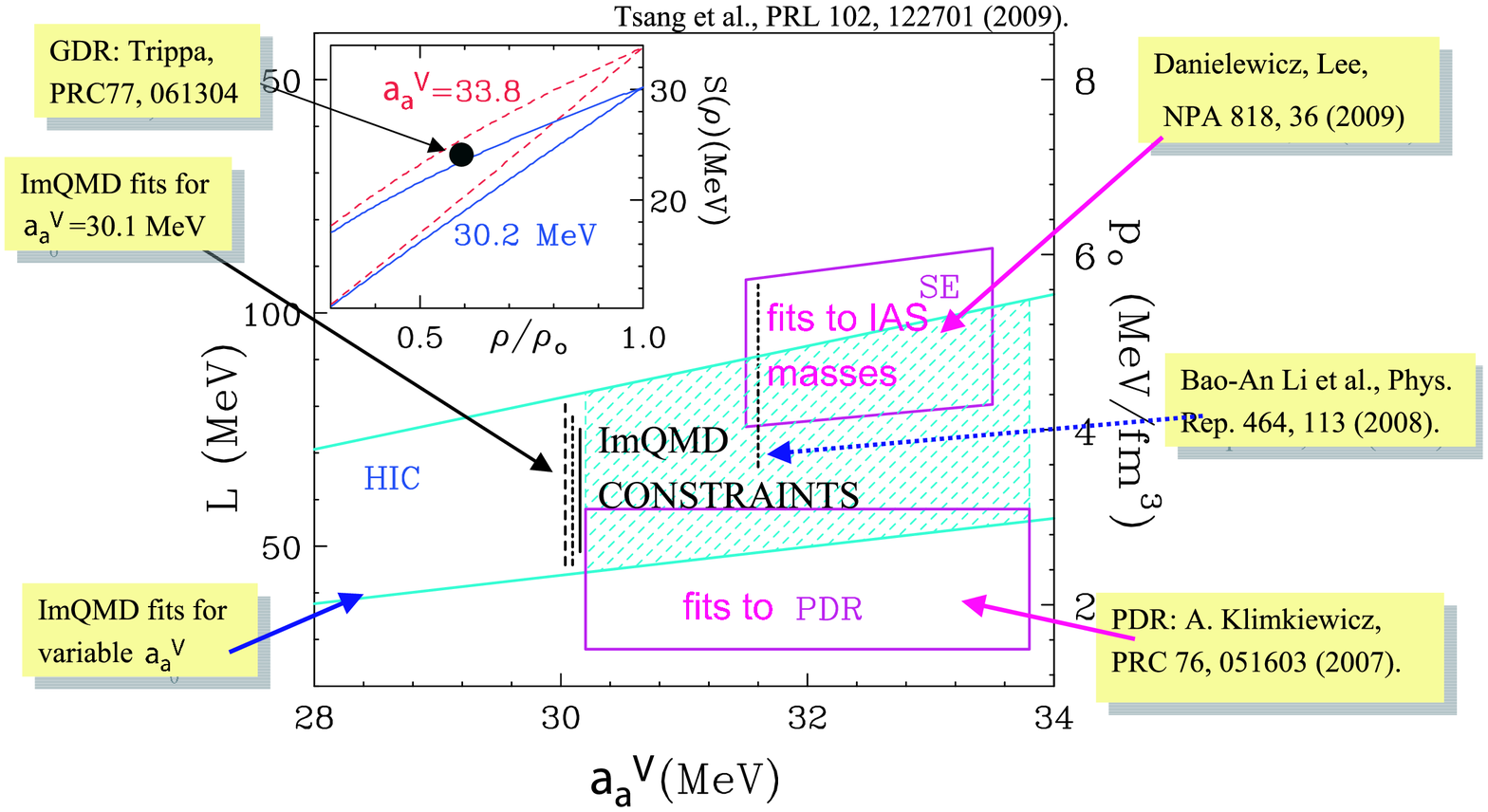}
\caption{Constraints on the parameters $a_a^V$ and $L$ of the symmetry energy, from different indicated sources, after Ref.~\cite{tsa09}.}
\label{fig:Betty09} 
\end{figure}

On the right of Fig.~\ref{fig:IR_bet09}, a comparison may be seen of the data~\cite{liu07} on the ratio $R$, from Sn + Sn collisions at 50 MeV/nucleon, to the results of calculations within the ImQMD model~\cite{zha08}.  In the calculations, symmetry energy of the form
\begin{equation}
S(\rho) = 12.3 \, \text{MeV} \, (\rho/\rho_0)^{2/3} + 17.6 \, \text{MeV} \, (\rho/\rho_0)^{\gamma_i} 
\end{equation}
has been used \cite{tsa09}.  The value of $\chi^2$, in the comparisons, minimizes in the vicinity of $\gamma_i \sim 0.7$. Implications for the symmetry energy from comparing central-reaction data to transport-model simulations are next shown in Fig.~\ref{fig:Betty09}.  Also results from the IAS analysis are shown there, together with results from analyzing collective excitations.  While there is some level of convergence for conclusions on the symmetry energy, reached from different directions, there is no consistency yet.  As has been mentioned, curvature corrections for teh IAS analysis are likely to lower both the $L$ and $a_a^V$ values.

To sum up, efforts to narrow down features of nuclear symmetry energy are advanced from different directions.  We have mostly discussed here an effort from the structure direction and a little that from the reaction side.  Qualitative understanding of connections between inputs and outcomes can be helpful in establishing the validity of the employed procedures as well as circumventing technical problems.  To bring consistency between results obtained though different methodologies, it is necessary to understand systematic errors associated with the methodologies.

This work was supported by the National Science Foundation under Grant No.\ PHY-0800026.


%
\end{document}